\documentclass[letter,twocolumn]{jpsj3}
\usepackage{txfonts}
\usepackage{color}
\usepackage{ulem}

\title{
Strong-Coupling Superconductivity in BaPd$_{2}$As$_{2}$ Induced by Soft Phonons in the ThCr$_{2}$Si$_{2}$-type Polymorph
}

\author{
Kazutaka Kudo\thanks{E-mail: kudo@science.okayama-u.ac.jp}, 
Yoshiaki Yamada, 
Takaaki Takeuchi, 
Takumi Kimura, 
Satoshi Ioka, 
Genta Matsuo, 
Yutaka Kitahama, 
and Minoru Nohara\thanks{E-mail: nohara@science.okayama-u.ac.jp}
}

\inst{
Research Institute for Interdisciplinary Science, Okayama University, Okayama 700-8530, Japan
}

\abst{
The specific heat of two polymorphs of BaPd$_{2}$As$_{2}$ was measured. 
The ThCr$_2$Si$_2$-type polymorph (space group $I4/mmm$, $D_{4h}^{17}$, No. 139) is a previously reported superconductor with a transition temperature $T_{\rm c}$ $\simeq$ 3.5 K, while the CeMg$_{2}$Si$_{2}$-type polymorph ($P4/mmm$, $D_{4h}^{1}$, No. 123) is a normal metal and does not exhibit superconductivity down to 1.8 K. 
Our results revealed that the ThCr$_2$Si$_2$-type has an anomalously low Debye temperature, indicative of soft phonons, compared to the CeMg$_{2}$Si$_{2}$-type. 
Moreover, a large specific-heat jump at $T_{\rm c}$ indicated that the superconductivity of ThCr$_2$Si$_2$-type is a strong-coupling type, which is likely derived from soft phonons. 
}

\begin{document}
\maketitle

Energetically low-lying phonons often result in strong-coupling superconductivity with an enhanced superconducting transition temperature $T_{\rm c}$ \cite{hiroi,nagao,oshiba,isono,guo,ishida,kudo1,hirai1,kudo4,hirai2}. 
Thus, engineering materials to produce low-lying phonons has become an important issue in superconductivity. 
A promising route to producing low-lying phonons is a local anharmonic vibration of the ion that is weakly bound in a cage-like structure. 
Such a vibration is called a rattling phonon and characterized by a low-frequency Einstein mode. 
A remarkable example is the $\beta$-pyrochlore oxide KOs$_{2}$O$_{6}$ with $T_{\rm c}$ = 9.6 K \cite{hiroi,nagao,oshiba,isono}, as well as 
Ba$_{3}$Ir$_{4}$Ge$_{16}$ with $T_{\rm c}$ $\simeq$ 6 K \cite{guo,ishida}. 
Another route to producing low-lying phonons is structural instability that becomes evident from the occurrence of a structural phase transition due to applied pressure or chemical doping. 
Prominent examples of such superconductors include the 122-type pnictides BaNi$_{2}$(As$_{1-x}$P$_{x}$)$_{2}$ with $T_{\rm c}$ = 3.3 K \cite{kudo1}, BaNi$_{2}$(Ge$_{1-x}$P$_{x}$)$_{2}$ with $T_{\rm c }$ = 2.9 K \cite{hirai1}, and Ba(Ni$_{1-x}$Cu$_{x}$)$_2$As$_2$ with $T_{\rm c}$ = 3.2 K \cite{kudo4}, all of which exhibit strong-coupling superconductivity because of a structural phase transition and the subsequent phonon softening characterized by an anomalously low Debye frequency. 
The MnP-type compound IrGe with $T_{\rm c}$ = 4.7 K is a rare example that exhibits strong-coupling superconductivity with low-lying phonons but does not exhibit a structural phase transition \cite{hirai2}. 
Such superconductors suggest how low-lying phonons may be produced for enhancing superconductivity.

The title compound BaPd$_{2}$As$_{2}$ has been known to crystallize in three forms: ThCr$_{2}$Si$_{2}$-type (space group $I4/mmm$, $D_{4h}^{17}$, No.~139) \cite{mewis1}, CeMg$_{2}$Si$_{2}$-type ($P4/mmm$, $D_{4h}^{1}$, No.~123)  \cite{mewis2}, and the intergrowth structure of CaBe$_{2}$Ge$_{2}$-type and CeMg$_{2}$Si$_{2}$-type ($I4/mmm$) \cite{mewis2}. 
The ThCr$_{2}$Si$_{2}$-type structure consists of PdAs$_{4}$ tetrahedra, while the CeMg$_{2}$Si$_{2}$-type consists of PdAs$_4$ planar squares, as shown in Figs.~1(a) and 1(b), respectively.
The intergrowth structure, shown in Fig.~1(c), consists of Pd$_{4}$As tetrahedra, which are characteristic of the CaBe$_{2}$Ge$_{2}$-type structure such as BaPt$_{2}$As$_{2}$ \cite{jiang,cyguo}, as well as PdAs$_{4}$ planar squares, which are characteristic of the CeMg$_{2}$Si$_{2}$-type structure. 
Guo et al. reported superconductivity at $T_{\rm c}$ = 3.85 K for the ThCr$_{2}$Si$_{2}$-type structure \cite{qi-guo}. 
This $T_{\rm c}$ of BaPd$_{2}$As$_{2}$ is anomalously high compared with $T_{\rm c}$ = 1.27 and 0.92 K for CaPd$_{2}$As$_{2}$ and SrPd$_{2}$As$_{2}$, respectively \cite{anand}, which are isoelectronic and isostructural with BaPd$_{2}$As$_{2}$ of the ThCr$_{2}$Si$_{2}$-type structure. 
On the other hand, no bulk superconductivity was observed for BaPd$_{2}$As$_{2}$ when the structure is the CeMg$_{2}$Si$_{2}$-type \cite{anand}. 
These observations suggest that a key factor in enhancing $T_{\rm c}$ may be present in BaPd$_{2}$As$_{2}$.

In this paper, we present the results of our synthesis of two polymorphs of BaPd$_{2}$As$_{2}$, namely, ThCr$_2$Si$_2$-type and CeMg$_{2}$Si$_{2}$-type, along with the results of specific-heat measurements. 
We found that when the structure was ThCr$_{2}$Si$_{2}$-type, BaPd$_{2}$As$_{2}$ exhibited superconductivity at $T_{\rm c}$ = 3.5 K with a large specific-heat jump at the superconducting transition $\Delta C/\gamma T_{\rm c}$ = 2.3, where $\gamma$ is the electronic specific-heat coefficient, which indicated that the superconductivity was a strong-coupling type. 
Further, low-lying phonons, which manifested themselves at an anomalously low Debye temperature $\Theta_{\rm D}$ = 144 K, was observed in the normal-state specific-heat data. 
On the other hand, when the structure was CeMg$_{2}$Si$_{2}$-type, BaPd$_{2}$As$_{2}$ did not exhibit superconductivity, and the Debye temperature of $\Theta_{\rm D}$ = 259 K was a typical value for 122-type pnictides.

\begin{figure}[t]
\begin{center}
\includegraphics[width=8.5cm]{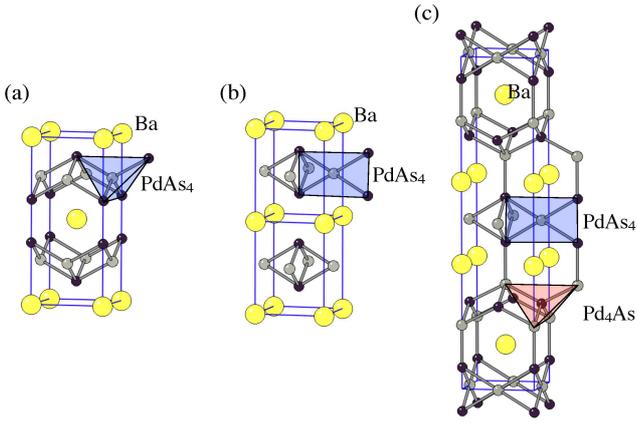}
\caption{
(Color online) The crystal structures of BaPd$_{2}$As$_{2}$ with (a) ThCr$_{2}$Si$_{2}$-type (space group $I4/mmm$, $D_{4h}^{17}$, No.~139) \cite{mewis1}, 
(b) CeMg$_{2}$Si$_{2}$-type ($P4/mmm$, $D_{4h}^{1}$, No.~123) \cite{mewis2}, 
and (c) the alternately stacked CaBe$_{2}$Ge$_{2}$-type and CeMg$_{2}$Si$_{2}$-type structure ($I4/mmm$, $D_{4h}^{17}$, No.~139) \cite{mewis2}. 
Blue solid lines indicate the unit cell. Two unit cells are shown for the CeMg$_{2}$Si$_{2}$-type in (b). 
}
\end{center}
\end{figure}

\begin{figure}[t]
\begin{center}
\includegraphics[width=7.5cm]{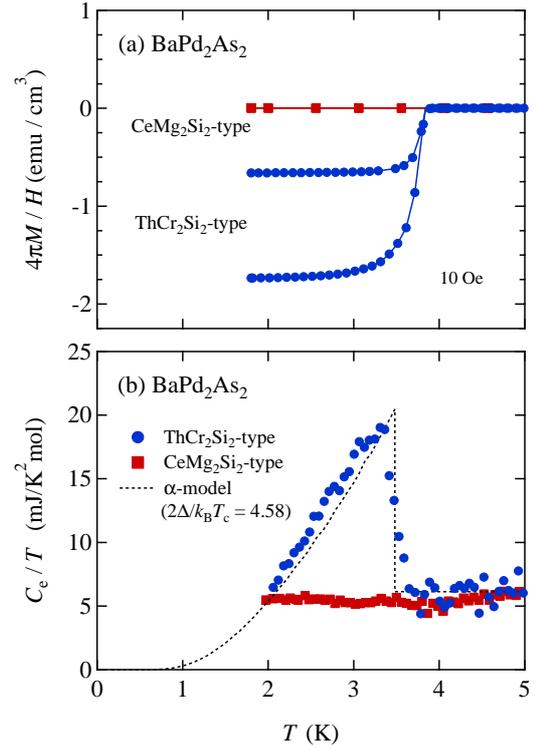}
\caption{
(Color online) (a) Temperature dependence of the dc magnetization $M$ measured in a magnetic field $H$ of 10 Oe for BaPa$_{2}$As$_{2}$ with ThCr$_{2}$Si$_{2}$-type and CeMg$_{2}$Si$_{2}$-type structures, under zero-field-cooling and field-cooling conditions.  
(b) Temperature dependence of the electronic specific heat divided by temperature, $C_{\rm e}/T$, for BaPa$_{2}$As$_{2}$ with ThCr$_{2}$Si$_{2}$-type and CeMg$_{2}$Si$_{2}$-type structures. $C_{\rm e}$ was determined from the total specific heat $C$, as shown in Fig.~3. 
The dashed curve denotes the temperature dependence of $C_{\rm e}/T$ calculated based on the $\alpha$-model \cite{alpha_model}, using $T_{\rm c}$ = 3.48 K, $\gamma$ = 6.12 mJ/K$^2$mol, and 2$\Delta/k_{\rm B}T_{\rm c}$ = 4.58, where $T_{\rm c}$  is the superconducting transition temperature, $\gamma$ is the electronic specific-heat coefficient, $\Delta$ is the superconducting gap at 0 K, and $k_{\rm B}$ is the Boltzmann constant.
}
\end{center}
\end{figure}

\begin{figure}[t]
\begin{center}
\includegraphics[width=7.5cm]{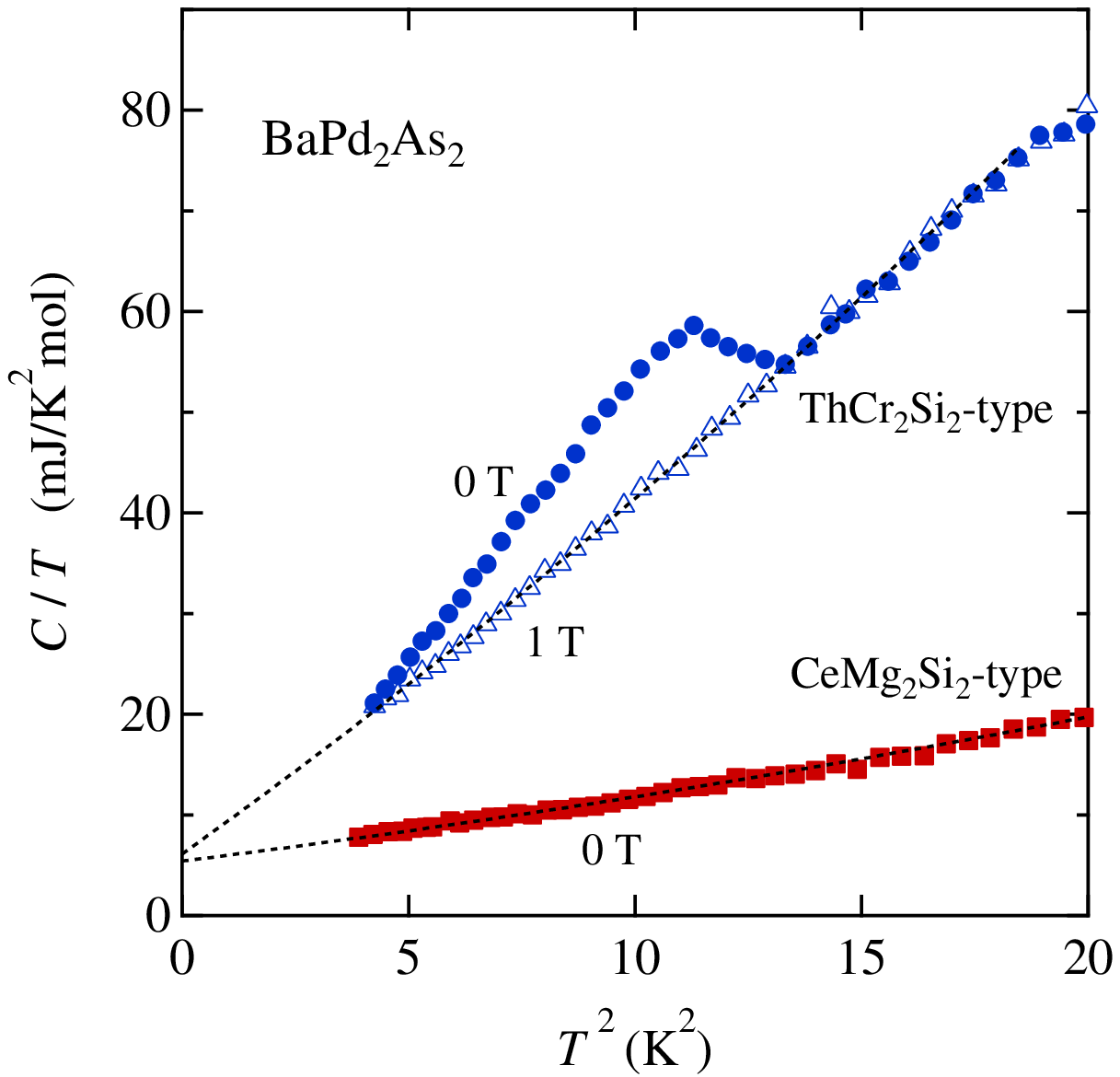}
\caption{
(Color online) The specific heat divided by the temperature, $C/T$, as a function of $T^{2}$ for BaPd$_{2}$As$_{2}$ with the ThCr$_{2}$Si$_{2}$-type and CeMg$_{2}$Si$_{2}$-type structures. 
The dashed curves denote the fits by $C/T = \gamma + \beta T^{2} + \delta T^{4}$, where $\gamma$ is the electronic specific-heat coefficient, $\beta$ is a constant corresponding to the Debye phonon contributions, and $\delta$ is a constant. 
}
\end{center}
\end{figure}

Each polymorph of BaPd$_{2}$As$_{2}$, ThCr$_{2}$Si$_{2}$-type or CeMg$_{2}$Si$_{2}$-type, was selectively synthesized with different conditions. 
Single crystals were obtained for both polymorphs. 
The condition for obtaining the ThCr$_{2}$Si$_{2}$-type was strict. 
First, a mixture of Pd$:$As = 1$:$1.5 was sealed in an evacuated quartz tube, heated at 500 $^{\circ}$C for 10 h, and then heated at 750 $^{\circ}$C for 40 h, followed by furnace cooling. 
The ampule was heated repeatedly at 800 and 850 $^{\circ}$C, following at 500 $^{\circ}$C in each process, without intermediate grinding. 
The resultant lumps of PdAs$_{2}$ with metallic luster were separated from the Pd$_{2}$As powder and pulverized. 
Second, stoichiometric amounts of the Ba, Pd, and PdAs$_{2}$ powders were placed in an alumina crucible, sealed in an evacuated quartz tube, and heated at 900 $^{\circ}$C for 48 h, followed by a quenching in iced water. 
As a result, single crystals of ThCr$_{2}$Si$_{2}$-type were obtained with a typical dimension of 0.1--0.5 $\times$ 0.1--0.5 $\times$ 0.02 mm$^{3}$. 
Note that a larger ratio of As in the starting composition resulted in As grains in the quartz tube during the PdAs$_2$ synthesis. 
In this case, a mixture of ThCr$_{2}$Si$_{2}$-type and CeMg$_{2}$Si$_{2}$-type crystals was obtained, even if the As grains were removed and only PdAs$_{2}$ lumps were used for the second step.  
On the other hand, the condition for obtaining the CeMg$_{2}$Si$_{2}$-type was less strict.
We used a precursor PdAs$_{2}$ synthesized from a mixture with Pd$:$As = 1$:$2. 
Stoichiometric amounts of Ba, Pd, and PdAs$_2$ were placed in an alumina crucible, sealed in an evacuated quartz tube, and heated at 700--1150 $^\circ$C, followed by either quenching in iced water or furnace cooling. 
Both conditions resulted in single crystals of CeMg$_{2}$Si$_{2}$-type with a typical dimension of 0.5 $\times$ 0.5 $\times$ 0.02 mm$^3$. 
All of these observations suggest that the CeMg$_{2}$Si$_{2}$-type is a stable phase, while the ThCr$_{2}$Si$_{2}$-type is a metastable phase, consistent with total energies predicted by first-principles calculations\cite{shein}.

Powder X-ray diffraction (XRD) studies confirmed that the resulting samples were a single phase of either ThCr$_{2}$Si$_{2}$-type or CeMg$_{2}$As$_{2}$-type. 
The magnetization $M$ and specific heat $C$ were measured using the Magnetic Property Measurement System (MPMS) and the Physical Property Measurement System (PPMS) by Quantum Design, respectively. 
The specific-heat measurement of a standard Cu sample yielded the electronic specific-heat coefficient $\gamma$ = 0.694(3) mJ/K$^{2}$mol and Debye temperature $\Theta_{\rm D}$ = 337.3(9) K, which agreed reasonably well with literature values \cite{Cu_std}.

The superconducting transition in BaPd$_{2}$As$_{2}$ was examined by magnetization measurements, as shown in Fig.~2(a). 
In a manner consistent with previous reports \cite{qi-guo,anand}, BaPd$_{2}$As$_{2}$ exhibited a superconducting transition at 3.85 K with the ThCr$_{2}$Si$_{2}$-type structure, while superconductivity was not observed down to 1.8 K (the lowest temperature measured) for the CeMg$_{2}$Si$_{2}$-type structure.

Figure 3 shows the specific heat divided by the temperature $C/T$ as a function of the squared temperature $T^{2}$. 
The normal-state data could be fitted by $C/T = \gamma + \beta T^{2} + \delta T^{4}$, where $\gamma$ is an electronic specific-heat coefficient, $\beta$ is the coefficient of phonon contributions from which the Debye temperature $\Theta_{\rm D}$ is estimated, and $\delta$ is a constant related to phonon contributions.  
As shown in Fig.~3, the slope of the $C/T$ vs. $T^{2}$ curve of the ThCr$_{2}$Si$_{2}$-type was significantly steeper than that of the CeMg$_{2}$Si$_{2}$-type, suggesting the presence of significantly soft phonons in the ThCr$_{2}$Si$_{2}$-type structure. 
We estimated the Debye temperature $\Theta_{\rm D}$ = 144 and 259 K for the ThCr$_{2}$Si$_{2}$-type and CeMg$_{2}$Si$_{2}$-type, respectively. 
In contrast to the remarkable difference in the Debye temperature, the electronic specific-heat coefficient $\gamma$ showed similar values for the two structures, as can be seen from the almost identical intercepts of the $C/T$ vs. $T^{2}$ curves along the $C/T$ axis in Fig.~3. 
The $\gamma$ values are summarized in Table~I. 
The observed $\gamma$ values were comparable to the first-principles calculation values, $\gamma$ = 5.664 and 4.639 mJ/K$^{2}$mol for the ThCr$_{2}$Si$_{2}$-type and CeMg$_{2}$Si$_{2}$-type structures, respectively \cite{shein}.

The superconducting transition temperature $T_{\rm c}$ = 3.5 K, determined by specific-heat measurements, as shown in Fig.~2(b), for BaPd$_{2}$As$_{2}$ was markedly higher than $T_{\rm c}$ = 1.27 and 0.92 K for CaPd$_{2}$As$_{2}$ and SrPd$_{2}$As$_{2}$, respectively \cite{anand}.  
Correspondingly, the Debye temperature $\Theta_{\rm D}$ = 144 K for BaPd$_{2}$As$_{2}$ was anomalously low compared with $\Theta_{\rm D}$ = 276 and 298 K for CaPd$_{2}$As$_{2}$ and SrPd$_{2}$As$_{2}$, respectively \cite{anand}. 
In contrast to the remarkable differences in both $T_{\rm c}$ and $\Theta_{\rm D}$ between BaPd$_{2}$As$_{2}$ and its isostructural compounds, CaPd$_{2}$As$_{2}$ and SrPd$_{2}$As$_{2}$, the electronic specific-heat coefficient $\gamma$ showed similar values, as summarized in Table~I. 
These observations suggested that the enhanced superconductivity in BaPd$_{2}$As$_{2}$ originated not from the electronic properties but from phonons, specifically, from enhanced electron--phonon coupling due to soft phonons. 
Indeed, the normalized specific-heat jump at the superconducting transition $\Delta C_{\rm e} / \gamma T_{\rm c}$ = 2.3, which was determined based on the $\alpha$-model \cite{alpha_model}, as shown in Fig.~2(b), indicated strong-coupling superconductivity due to the enhanced electron--phonon coupling in BaPd$_{2}$As$_{2}$. On the other hand, the small value of $\Delta C_{\rm e} / \gamma T_{\rm c}$ $\simeq$ 1 indicated weak-coupling superconductivity in CaPd$_{2}$As$_{2}$ and SrPd$_{2}$As$_{2}$ \cite{anand}.

\begin{figure}[t]
\begin{center}
\includegraphics[width=7.5cm]{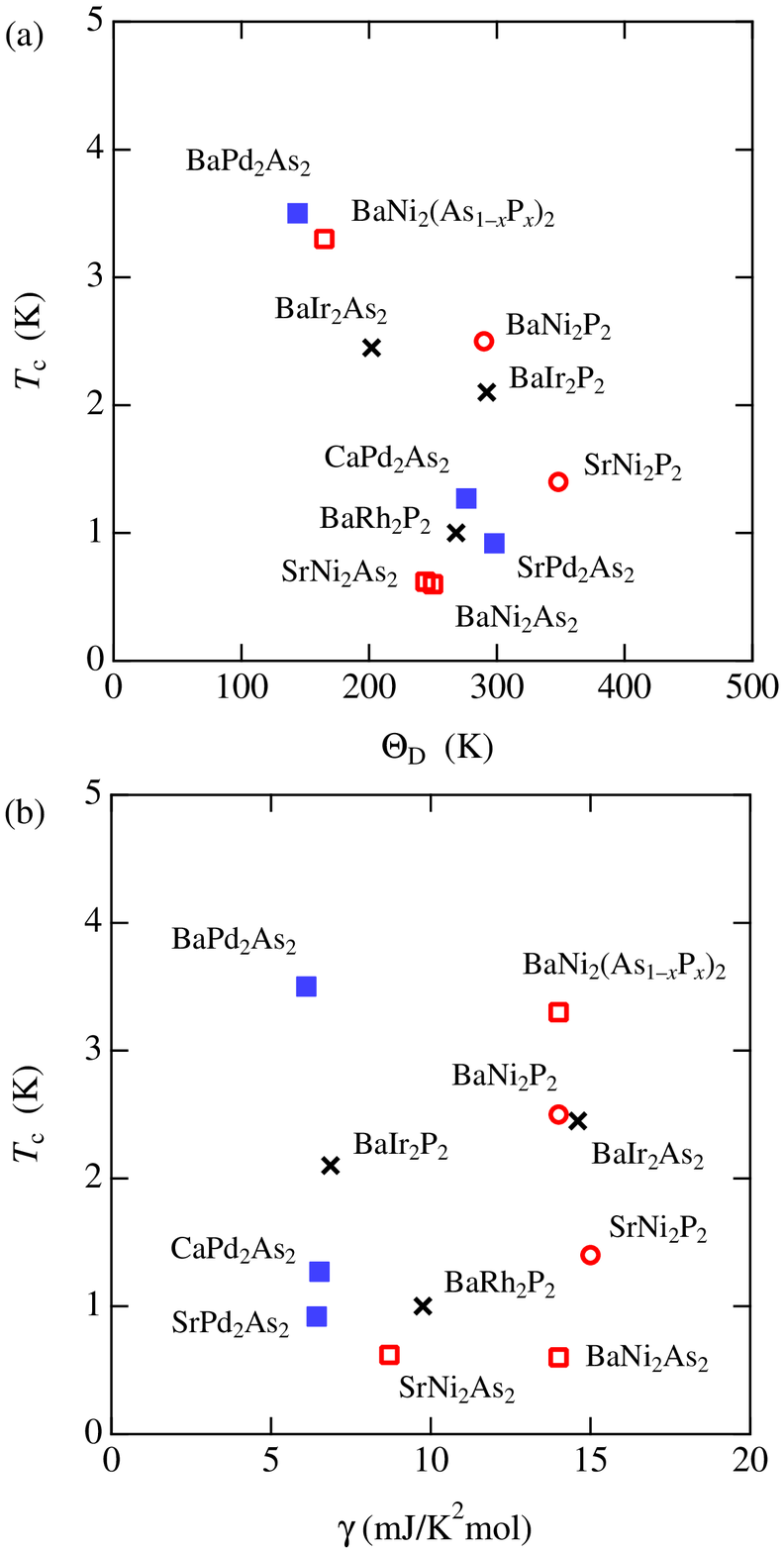}
\caption{
(Color online) (a) The superconducting transition temperature $T_{\rm c}$ vs. Debye temperature $\Theta_{\rm D}$, and (b) $T_{\rm c}$ vs. electronic specific-heat coefficient $\gamma$ for AETM$_{2}$X$_{2}$ (AE = Ca, Sr, and Ba; TM = Rh, Ir, Ni, and Pd; X = P and As) with the tetragonal ThCr$_{2}$Si$_{2}$-type structure:  
SrNi$_{2}$As$_{2}$ \cite{bauer}, 
BaNi$_{2}$P$_{2}$ \cite{hirai1}, 
BaNi$_{2}$(As$_{1-x}$P$_{x}$)$_{2}$ with $x$ = 0.077 \cite{kudo1}, 
CaPd$_{2}$As$_{2}$ \cite{anand}, 
SrPd$_{2}$As$_{2}$ \cite{anand}, 
BaPd$_{2}$As$_{2}$ (this work), 
BaRh$_{2}$P$_{2}$ \cite{hirai3}, 
BaIr$_{2}$P$_{2}$ \cite{hirai3}, 
BaIr$_{2}$As$_{2}$ \cite{Wang}; 
and the distorted ThCr$_{2}$Si$_{2}$-type structure: 
SrNi$_{2}$P$_{2}$ \cite{ronning} and 
BaNi$_{2}$As$_{2}$ \cite{kudo1}. 
}
\end{center}
\end{figure}

\begin{table}[t]
\caption{
The superconducting transition temperature $T_{\rm c}$, electronic specific-heat coefficient $\gamma$, Debye temperature $\Theta_{\rm D}$, and normalized specific-heat jump $\Delta C_{\rm e}/\gamma T_{\rm c}$ for AEPd$_{2}$As$_{2}$ (AE = Ca, Sr, and Ba) with the ThCr$_{2}$Si$_{2}$-type structure  (space group $I4/mmm$, $D_{4h}^{17}$, No.~139) and for BaPd$_{2}$As$_{2}$ with the CeMg$_{2}$Si$_{2}$-type structure ($P4/mmm$, $D_{4h}^{1}$, No.~123). 
}
\begin{center}
\begin{tabular}{llcccc}
\hline
Compound & Structure type&  $T_{\rm c}$ &  $\gamma$ &  $\Theta_{\rm D}$ & $\Delta C_{\rm e}/\gamma T_{\rm c}$\\
 &  &   (K) &  (mJ/K$^{2}$mol) &  (K)&  \\
\hline
CaPd$_{2}$As$_{2}$* & ThCr$_{2}$Si$_{2}$ & 1.27 & 6.52 & 276 & 1.1\\
SrPd$_{2}$As$_{2}$* & ThCr$_{2}$Si$_{2}$  & 0.92 & 6.43 & 298 & 1.0\\
BaPd$_{2}$As$_{2}$   & ThCr$_{2}$Si$_{2}$ & 3.5 & 6.1 & 144 & 2.3\\ 
BaPd$_{2}$As$_{2}$   & CeMg$_{2}$Si$_{2}$ & -- & 5.4 & 259 & -- \\ 
\hline
* Ref.~\citen{anand}.
\end{tabular}
\end{center}
\end{table}

The anomalously low Debye temperature $\Theta_{\rm D}$ and enhanced superconducting transition temperature $T_{\rm c}$ in BaPd$_{2}$As$_{2}$ was visualized using a $T_{\rm c}$ vs. $\Theta_{\rm D}$ plot for various 122-type compounds AETM$_{2}$X$_{2}$, where AE = Ca, Sr, and Ba; TM = Rh, Ir, Ni, and Pd; and X = P and As, with the tetragonal ThCr$_{2}$Si$_{2}$-type and distorted ThCr$_{2}$Si$_{2}$-type structures, as shown in Fig.~4(a). 
From the figure, lower values of $\Theta_{\rm D}$ tend to have higher values of $T_{\rm c}$; note that BaPd$_{2}$As$_{2}$ is placed at the highest value of $T_{\rm c}$ and lowest value of $\Theta_{\rm D}$ among the family of 122-type compounds with the ThCr$_{2}$Si$_{2}$-type structure, except for iron-based superconductors (not shown) such as BaFe$_{2}$As$_{2}$, in which the mechanism of superconductivity is different \cite{Fe_ishida,Fe_hosono}. 
This trend was evident in the series of CaPd$_{2}$As$_{2}$--SrPd$_{2}$As$_{2}$--BaPd$_{2}$As$_{2}$. 
However, this trend was less apparent in SrNi$_{2}$P$_{2}$--BaNi$_{2}$P$_{2}$ and SrNi$_{2}$As$_{2}$--BaNi$_{2}$As$_{2}$ likely as a result of the structural phase transition and resultant structural distortion of SrNi$_{2}$P$_{2}$ \cite{ronning,kudo3} and BaNi$_{2}$As$_{2}$ \cite{kudo1,sefat}. 
However, the substitution of tiny amount of P for As in BaNi$_{2}$As$_{2}$ resulted in an enhanced $T_{\rm c}$ with lowered $\Theta_{\rm D}$ \cite{kudo1}, which was accompanied by the suppression of the structural phase transition \cite{kudo1}, and the $T_{\rm c}$--$\Theta_{\rm D}$ relation became evident in SrNi$_{2}$As$_{2}$--BaNi$_{2}$(As$_{1-x}$P$_{x}$)$_{2}$ in Fig.~4(a). 
On the other hand, in the $T_{\rm c}$ vs. $\gamma$ plot, shown in Fig.~4(b), $T_{\rm c}$ is not obviously dependent on the electronic specific-heat coefficient $\gamma$. Moreover, CaPd$_{2}$As$_{2}$--SrPd$_{2}$As$_{2}$--BaPd$_{2}$As$_{2}$ exhibited different $T_{\rm c}$ values, even though they exhibited similar $\gamma$-values (approximately 6 mJ/K$^{2}$mol). 
Similarly, the Ni-based compounds also exhibited quite different $T_{\rm c}$ values, even though their $\gamma$ values were similar (approximately 14 mJ/K$^{2}$mol).

All of these observations indicated that the determining factor of $T_{\rm c}$ was the soft phonons, or energetically low-lying phonons, of the ThCr$_{2}$Si$_{2}$-type structure. 
For BaNi$_{2}$As$_{2}$, the in-plane vibrations of Ni and As have been predicted to produce low-lying phonons \cite{subedi}, which eventually resulted in a structural phase transition that distorted the Ni square network into zigzag chains \cite{sefat}, thus suggesting freezing of the in-plane Ni motion. 
On the other hand, BaPd$_{2}$As$_{2}$ did not exhibit a structural phase transition in spite of the enhanced electron--phonon coupling. 
The lack of a structural phase transition could explain why BaPd$_{2}$As$_{2}$ exhibited the highest $T_{\rm c}$ value among AETM$_{2}$X$_{2}$. 
The phonon mode responsible for the lowered Debye temperature $\Theta_{\rm D}$ of BaPd$_{2}$As$_{2}$ was not clear from this study. 
We speculate that the in-plane Pd and As vibrations have importance, according to the analogy with BaNi$_{2}$As$_{2}$, as well as the first-principles calculations \cite{shein,nekrasov,chen}. 
We suggest the in-plane As vibration has particular importance to low-lying phonons because the ThCr$_{2}$Si$_{2}$-type structure could be transformed to the CeMg$_{2}$Si$_{2}$-type structure by the displacement of As, and vice versa, as can be seen from Figs. 1(a) and 1(b).
Moreover, the ThCr$_{2}$Si$_{2}$-type is a metastable phase, while the CeMg$_{2}$Si$_{2}$-type is a stable phase, as we mentioned above. 
Thus, we expect that the ThCr$_{2}$Si$_{2}$-type polymorph possesses low-lying phonons due to the structural instability of metastable phase. 
If this is the case, polymorphism could provide another route to producing low-lying phonons and hence enhanced superconductivity \cite{xie,pyon}.

In conclusion, we have measured the specific heat of BaPd$_{2}$As$_{2}$ with two different forms of BaPd$_{2}$As$_{2}$: the superconducting ThCr$_{2}$Si$_{2}$-type (space group $I4/mmm$, $D_{4h}^{17}$, No.~139) and non-superconducting CeMg$_{2}$Si$_{2}$-type ($P4/mmm$, $D_{4h}^{1}$, No.~123). 
The very large specific-heat jump at $T_{\rm c}$ indicated that the ThCr$_{2}$Si$_{2}$-type was a strong-coupling superconductor. 
Compared with the CeMg$_{2}$Si$_{2}$-type and other ThCr$_{2}$Si$_{2}$-type pnictides, an anomalously low Debye temperature was observed in ThCr$_{2}$Si$_{2}$-type BaPd$_{2}$As$_{2}$, originating from low-lying phonons. 
The key factor that enhanced $T_{\rm c}$ in BaPd$_{2}$As$_{2}$ was low-lying phonons with a Debye temperature of $\Theta_{\rm D}$ = 144 K of the ThCr$_{2}$Si$_{2}$-type polymorph.

\acknowledgment
The authors are grateful to M. Akoshima and H. Abe for helping to calibrate the instrument for specific-heat measurements, and to K. Fujimura, N. Nishimoto, T. Mizukami, and H. Ishii for their technical assistance in the experiments and data analyses for this work. 
A part of this work was performed at the Advanced Science Research Center, Okayama University.
This work was partially supported by Grants-in-Aid for Scientific Research (No. 26287082, 15H01047, 15H05886, and 16K05451) provided by the Japan Society for the Promotion of Science (JSPS) 
and the Program for Advancing Strategic International Networks to Accelerate the Circulation of Talented Researchers from JSPS.

\end{document}